%% file: main.tex
\documentclass[journal,twocolumn]{IEEEtran}

\usepackage{stfloats}

\usepackage{graphicx}
\usepackage{textcomp}
\usepackage{xcolor}
\usepackage{caption}
\usepackage{multicol,mwe,float,subcaption}
\usepackage{makecell}
\usepackage{float}

\usepackage[T1]{fontenc}
\usepackage{cite}
\usepackage{amsmath,amssymb}
\usepackage{bm}
\usepackage{graphicx}
\usepackage{textcomp}
\usepackage{xcolor}
\usepackage{threeparttable}
\usepackage{multirow}
\usepackage{verbatim}
\usepackage{url}
\usepackage{hyperref}
\usepackage{comment}
\usepackage{color}
\usepackage{booktabs}
\usepackage{amsthm}
\usepackage{makecell}
\usepackage{cuted}
\usepackage{orcidlink}

\usepackage[linesnumbered, ruled]{algorithm2e}
\SetKwRepeat{Do}{do}{while}%

\newtheorem{lemma}{{Lemma}} 
\newtheorem{proposition}{Proposition}

\newcommand{\rank}{\operatorname{rank}}

\makeatletter
\newcommand{\ie}{\emph{i.e.}\@ifnextchar.{\!\@gobble}{}}
\newcommand{\eg}{\emph{e.g.}\@ifnextchar.{\!\@gobble}{}}
\newcommand{\etc}{etc\@ifnextchar.{}{.\@}}
\makeatother
\usepackage[font=small]{caption}
\begin{document}

\title{Robust Hybrid Precoding for Millimeter Wave MU-MISO System Via Meta-Learning}

\author{
{Yifan Guo~\orcidlink{0009-0004-1393-8839}}
}
\vspace{-10pt}


\maketitle

\pagestyle{empty}  
\thispagestyle{empty} 

\begin{abstract}
Thanks to the low cost and power consumption, hybrid analog-digital architectures are considered as a promising energy-efficient solution for massive multiple-input multiple-output (MIMO) systems. The key idea is to connect one RF chain to multiple antennas through low-cost phase shifters. However, due to the non-convex objective function and constraints, we propose a gradient-guided meta-learning (GGML) based alternating optimization framework to solve this challenging problem. The GGML based hybrid precoding framework is \textit{free-of-training} and \textit{plug-and-play}. Specifically, GGML feeds the raw gradient information into a neural network, leveraging gradient descent to alternately optimize sub-problems from a local perspective, while a lightweight neural network embedded within the meta-learning framework is updated from a global perspective. We also extend the proposed framework to include precoding with imperfect channel state information. Simulation results demonstrate that GGML can significantly enhance spectral efficiency, and speed up the convergence by 8 times faster compared to traditional approaches. Moreover, GGML could even outperform fully digital weighted minimum mean square error (WMMSE) precoding with the same number of antennas.
\end{abstract}


\input{introduction}

\input{system}

\input{method}

\input{imcsi}

\input{simulation}

\input{conclusion} 

\input{appendix}

\small
\bibliographystyle{IEEEtran}
\bibliography{reference}
\vspace{12pt}

\end{document}

%% file: introduction.tex
\section{Introduction}\label{sec:intro}
\IEEEPARstart{T}{he} millimeter wave (mmWave) communication is considered a highly promising technology for future wireless communication due to its abundant spectrum resources and its ability to provide gigabit-level data rates\cite{mmWave1, mmWave2, mmWave3, mmWave4}. However, mmWave communication still faces challenges such as significant propagation losses and limited coverage, necessitating the deployment of large-scale antenna arrays to achieve beamforming gains. In contrast to conventional precoding schemes, mmWave systems implement hybrid analog-digital beamforming architectures, specifically designed to reduce hardware complexity and minimize power consumption\cite{hybridbeam1, hybridbeam2}. This architecture uses a limited number of RF chains connected to antennas via phase shifters, and the joint design of RF analog precoding and baseband digital precoding is required. However, hybrid precoding design in such architectures is a challenging task due to coupled variables and the unit modulus constraints imposed by phase shifters.
\par
\subsection{Related Works}
Early research primarily delved into hybrid precoding in single-user multiple-input multiple-output (SU-MIMO) scenarios. 
In \cite{omp}, the authors exploited the sparse scattering characteristics of mmWave channels to reformulate the original problem as a matrix reconstruction problem under sparse constraints and solved it using the orthogonal matching pursuit (OMP) algorithm. Alternating least squares \cite{yuweihybrid} and manifold optimization (MO) \cite{yxhhybrid} were also proposed for hybrid precoding, alternately optimizing by relaxing constant modulus constraints. An alternative approach in \cite{fshp} established a functional relationship between the precoding matrix and the analog precoding matrix, simplifying the original problem. In \cite{tianlinmmse}, the mean squared error (MSE) was employed as a performance metric to characterize transmission reliability, and hybrid precoding matrices were derived using generalized eigenvalue decomposition.
\par
Extending hybrid precoding to multi-user scenarios is more aligned with the future trends of wireless communications. For multi-user multiple-input single-output (MU-MISO) systems, the authors in \cite{yuweihybrid} divided the precoder design into two stages: element-wise optimization for the analog precoder, followed by the digital precoder using the zero-forcing algorithm. To optimize the total MSE of the system, truncated singular value decomposition was used to determine the analog precoder, while a closed-form solution was derived for updating the digital precoder \cite{trace}. Additionally, to balance performance and complexity, an equivalent gain transmission approach was used to update the analog precoder \cite{lelianghybrid}. For more complex MU-MIMO systems, if the number of RF chains is twice the number of data streams, hybrid beamforming architectures can reach the upper bound performance \cite{yuweihybrid}. Other research has explored hybrid precoding with fewer RF chains. In \cite{zxfhybrid}, the weighted minimum mean square error (WMMSE) algorithm \cite{WMMSE} and MO were used to solve the MMSE combiner and analog combiner, respectively. In \cite{wmmse-omp}, a framework based on WMMSE and OMP was used to obtain a locally optimal digital precoder. However, these methods are hindered in practical deployment in dynamic wireless networks due to performance degradation and high complexity.
\par
Recent years have witnessed increasing integration of deep learning (DL) into physical layer communications due to its outstanding performance and low computational complexity, including channel estimation \cite{hhtchannel_estimation, mwychannel_estimation, mxschannel_estimation_feedback}, channel state information (CSI) feedback \cite{mxschannel_estimation_feedback, wckCSI_feedback}, and signal detection \cite{hhtsignal_detection, Adaptive_signal_detection}. The main advantage of DL is that with abundant training samples, complex computational tasks can be shifted to the offline training phase, only simple forward computation is required during the online prediction phase. In hybrid precoding design, DL has been utilized to approximate the mapping between the inputs of traditional optimization algorithms and their corresponding optimal solutions, thereby circumventing the cumbersome iterative solving process and effectively managing non-convex optimization problems.This methodology has been extensively explored in both single-user \cite{HuangHongji, tianlin, huqiyu} and multi-user \cite{mwychannel_estimation,hybrid_multi_user1,hybrid_GAN_multi_user2,jhhybrid} scenarios. However, data-driven black-box neural networks (NNs) have drawbacks in terms of interpretability and generalization, making them susceptible to environmental changes. Model-driven DL, which integrates expert knowledge into network design, has garnered considerable attention in precoding tasks due to its robustness and interpretability. In \cite{model-driven-MO}, the authors embedded the step size as a trainable parameter into the MO process to enhance performance. Similarly, \cite{unfold-PGA} used data to learn iteration-related hyperparameters with predefined iteration counts to unfold the projected gradient ascent algorithm. In \cite{unfold-Mo-AltMin}, a sparse connection network was used to replace the complex steps of manifold optimization, and the tanh activation function was employed to ensure the constant modulus constraint, offering higher performance and lower complexity compared to the unfolding projected gradient ascent algorithm.
\par
To further enhance the robustness of DL in dynamic scenarios, researchers have explored precoding strategies based on transfer learning \cite{transfer_learning} and meta learning \cite{meta_learning1,meta_learning_sd}, demonstrating strong adaptability in such environments. However, the aforementioned learning schemes often require high-quality training and adaptation data, making their deployment impractical. To reduce the overhead of pretraining, \cite{zfh_meta,xjy_meta,yzx_meta} proposed a novel meta learning based optimization framework, enabling the algorithm to escape local optima and achieve improved performance. To the best of our knowledge, the design of hybrid precoding optimization using meta-learning in millimeter-wave systems remains largely unexplored, although it is an emerging trend.
\par
Most of the aforementioned studies assume the availability of perfect CSI, which is not always feasible. In \cite{zhangmaojun}, synthetic noise was introduced into the training dataset to make the network robust against channel errors, although the performance improvements from this training strategy were limited. To address the challenges of channel estimation related to hybrid architectures, \cite{model-driven-MO} proposed a bounded CSI error model and treated the channel error as equivalent to received noise, using DL to learn the corresponding mapping. In \cite{unfold-PGA}, the projected conceptual mirror prox minimax optimizer was employed to optimize the worst-case weighted sum rate maximization problem, where the step size of optimizer was a trainable parameter. Additionally, \cite{laingkai} addressed the outage constraint problem caused by imperfect channels using a model- and data-driven DL approach. However, the performance of these works could be further improved with meta-learning-assisted optimization algorithms, which motivates the design of corresponding algorithms and their extension to imperfect CSI scenarios.
\par
\subsection{Contributions and Organization}
In this study, we delve into the design of a hybrid precoder in mmWave MU-MISO systems. We propose an algorithm that integrates meta-learning principles with an iterative framework, facilitating a training-agnostic and ready-to-deploy approach\footnote{Simulation codes are provided to reproduce the results presented in this article: https://github.com/mistybeep/GGML.}. Furthermore, we incorporate the Kolmogorov–Arnold Networks (KAN) model as an embedded network to enhance performance. Additionally, we extend the meta-learning-assisted optimization method to account for imperfect CSI scenarios. The key contributions of this study are as follows:           
\begin{itemize}
    \item \emph{Development of a learning-assisted iterative algorithm for mmWave hybrid architectures:} We propose a gradient-guided meta-learning (GGML) approach that is plug-and-play and free of pre-training. Unlike DL, which directly learns the mapping from CSI to the precoding matrix, we use the gradient of the precoding matrix as input. Local optimization of coupled variables is performed from a local perspective, while global optimization of embedded network parameters is conducted from a global perspective, addressing the complexity of hybrid precoding. To further enhance algorithm performance, we incorporate KAN to capture key information from the input gradient information. The integration of meta-learning and alternating optimization results in a high-performance, low-complexity framework.
    \item \emph{Enhancement for imperfect CSI scenarios:} Recognizing the challenges of channel estimation with hybrid architecture, we propose a worst-case hybrid precoding design problem. The objective is to maximize the worst-case weighted spectral efficienc (SE) during transmission, which can be reformulated as a geometric program (GP) involving the geometric MSE. By incorporating GGML, we have enhanced its precoding performance in scenarios with imperfect CSI.
    \item \emph{Validation through comprehensive simulations:} Simulation results indicate that the GGML-based hybrid precoding framework can effectively improve system performance while accelerating convergence. Notably, the proposed GGML-based framework also can serve as a general solution for hybrid precoding in MU-MIMO systems. Additionally, the extended framework exhibits robustness and shows promise for practical deployment in scenarios with imperfect CSI.
\end{itemize}
\par
The remainder of the paper is structured as follows: Section \ref{sec:sys} introduces the system and channel model, followed by the formulation of optimization problems for both perfect and imperfect CSI scenarios. In Section \ref{sec:ggml}, we propose the GGML algorithm to solve the optimization problem under perfect CSI. In Section \ref{sec:imcsi}, we transform the optimization problem under imperfect CSI and extend our method to include imperfect CSI scenarios. Section \ref{sec:simulation} provides experimental details and numerical results. Finally, our work concludes in Section VI.
\par
\subsection{Notations}
The fonts $a$, $\mathbf{a}$, and $\mathbf{A}$ represent a scalar, a vector, and a matrix, respectively. $(\mathbf{A})_{i,j}$ is the element in the $i$-th row and $j$-th column of matrix $\mathbf{A}$. The conjugate, transpose, conjugate transpose, and inverse of $\mathbf{A}$ are denoted by $\mathbf{A}^\ast$, $\mathbf{A}^T$, $\mathbf{A}^H$, and $\mathbf{A}^{-1}$, respectively. $\|\mathbf{A}\|$ represents the Frobenius norm of $\mathbf{A}$. For symmetric matrices $\mathbf{A}$ and $\mathbf{B}$, $\mathbf{A} \succeq \mathbf{B}$ indicates that $\mathbf{A} - \mathbf{B}$ is positive semi-definite. $\mathbf{I}_{L}$ represents $L\times L$ identity matrix. $\circ$ denotes the composition of two mappings. ${[\cdot]^{+}}$ denotes the negative truncation operation, \ie, ${[x]^{+}}=\max\{x,0\}$.

%% file: system.tex
\vspace{2mm}
\section{System Model And Problem Formulation}\label{sec:sys}
In this section, we begin by presenting an overview of the MU-MISO system model with a hybrid structure. Subsequently, we formulate the hybrid precoding problem for both perfect and imperfect CSI scenarios.
\subsection{System Model}
\begin{figure}[t]\vspace{-0mm}
	\begin{center}		\centerline{\includegraphics[width=0.5\textwidth]{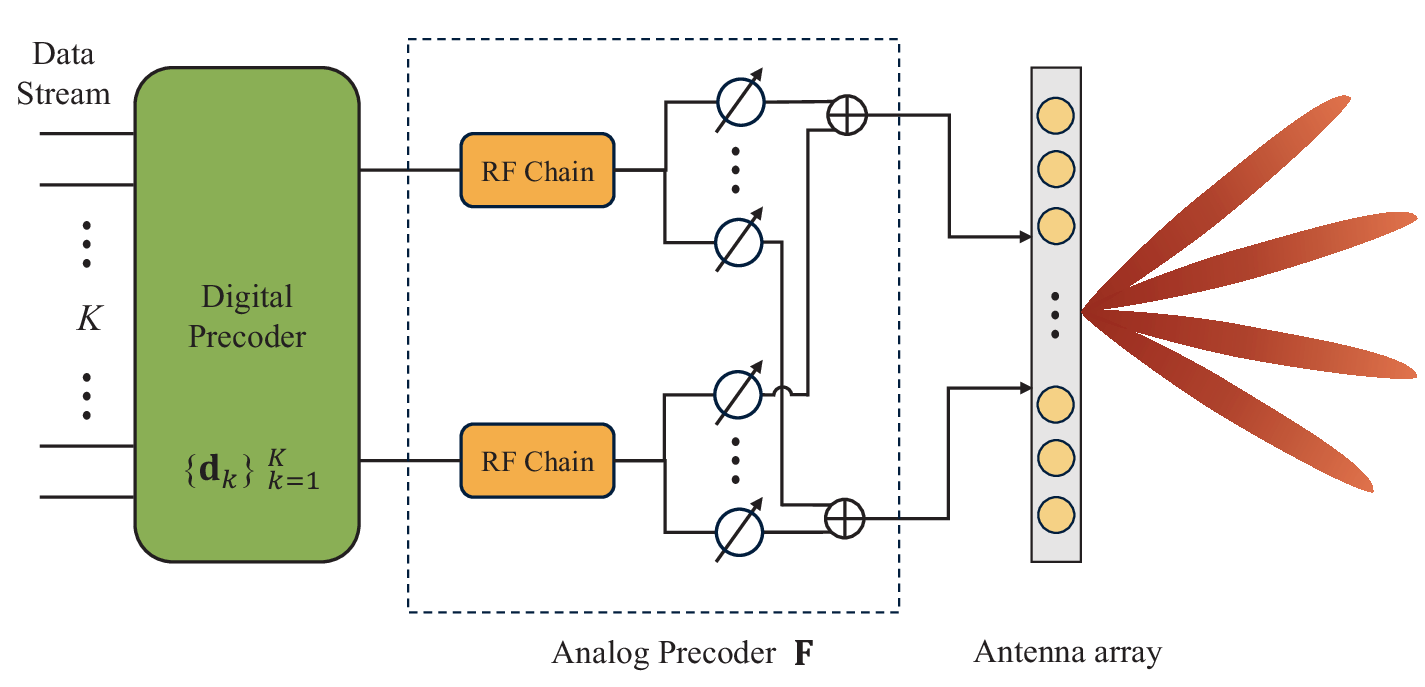}}  \vspace{-0mm}
	    \captionsetup{font=footnotesize, name={Fig.}, labelsep=period}  
        \caption[t]{\raggedright The architecture of hybrid precoding.}
		\label{fig:system} \vspace{-10mm}
	\end{center}
\end{figure}

As shown in Fig \ref{fig:system}, we consider the downlink transmission of the mmWave MU-MISO system. The BS is equipped with $N$ antennas to serve $K$ single-antenna users simultaneously. To reduce energy consumption and hardware costs, the transmitter employs a hybrid architecture, equipped with $M$ transmit RF chains, each RF chain connected to all antennas through phase shifters.
\par
At the transmitter, a digital transmit precoder $\mathbf{d}_k \in \mathbb{C}^{N}$ is first employed to precode the data stream of the $k$-th user, denoted as $s_k$ and assumed to be independent with zero mean and unit variance. Then, each stream is up-converted to the carrier frequency by passing through a dedicated RF chain. Before transmitting the RF signals at the $N$ antennas, an analog precoder composed of a number of phase shifters is deployed to enhance the array gain. From the perspective of the equivalent baseband, the transmitted signal vector sent to the $k$-th user is represented by $x_k=\mathbf{F}\mathbf{d}_ks_k$, where $\mathbf{F}\in \mathbb{C}^{N\times M}$ denotes the analog precoder shared by all users. Due to the hardware limitations of the phase shifters, the analog precoder is modeled as a matrix with entries of unit magnitude, namely, 
\begin{equation}
    \label{unit_magnitude}
    \mathcal{F}=\Big\{\mathbf{F}\in\mathbb{C}^{N\times M}\Big|\big|[\mathbf{F}]_{i,j}\big|=1,\quad\forall(i,j)\Big\}.
\end{equation}

The mmWave channel from the BS to user $k$ is denoted as $\mathbf{h}_k$ and is modeled using the Saleh-Valenzuela channel model, which is given by 
\begin{equation}
\label{channel}
\mathbf{h}_{k}=\sqrt{\frac{N}{N_{\mathrm{c}}N_{\mathrm{ray}}}}\sum_{c=1}^{N_{c}}\sum_{l=1}^{N_{\mathrm{ray}}}h_{cl}^{k}\mathbf{a}_{\mathrm{t}}(\phi_{cl}^{k})^{{H}},
\end{equation}
where $N_{\mathrm{c}}$ and $N_{\mathrm{ray}}$ denote the number of clusters and the number of rays in each cluster, respectively. The complex channel gain and the angles of departure of the $l$-th ray in the $c$-th scattering cluster are denoted as $h_{cl}$ and $\phi_{cl}^{k}$, respectively. In addition, for a uniform linear array (ULA) with antenna spacing of half a wavelength, the array response vector $\mathbf{a}_{\mathrm{t}}(\phi)$ can be represented as
\begin{equation}
\label{array_response}
\mathbf{a}_{\mathrm{t}}\big(\phi\big) = \frac{1}{\sqrt{N}} \Big[1, e^{\mathrm{j} \pi \sin(\phi)}, \ldots, e^{\mathrm{j} \pi (N-1) \sin(\phi)}\Big]^T.
\end{equation}

Assuming all channels are quasi-static and known, the equivalent baseband received signal at user $k$ can be represented as:
\begin{equation}
\label{signal}
y_k = \mathbf{h}_{k} \mathbf{F} \mathbf{d}_k s_k + \sum_{m \neq k,m=1}^{K}\mathbf{h}_{k}\mathbf{F} \mathbf{d}_m s_m + n_k,
\end{equation}
where $n_k \sim \mathcal{CN}(0,\sigma_{k}^2)$ denotes the additive white Gaussian noise at user $k$. In this expression, $\mathbf{F} \mathbf{d}_k s_k$ represents the desired signal at the $k$-th user, and other user signals are treated as interference to user $k$.
\par
The SE of the system is given by
\begin{equation}
    \label{system_rate}
    \mathcal{R}=\sum_{k=1}^{K} \alpha_k \mathcal{R}_k,
\end{equation}
where $\alpha_k$ is the priority of user $k$. $\mathcal{R}_k$ denotes the achievable spectral efficiency of the BS-to-user $k$ channel, which is given by
\begin{equation}
\label{rate}
\mathcal{R}_k = \log\Bigg(1+\frac{|\mathbf{h}_{k} \mathbf{F} \mathbf{d}_k|^2}{\sum_{i=1,i\neq k}^{K}|\mathbf{h}_{k}\mathbf{F} \mathbf{d}_i|^2 + \sigma_{k}^2}\Bigg).
\end{equation}
\subsection{Problem Formulation}
     
\subsubsection{Perfect CSI}We first address the hybrid precoding problem in scenarios with perfect CSI, aiming to maximize the WSR of the system. The problem can be formulated as
\begin{subequations}
\label{Optimization_Problem1}
    \begin{align}
    (\textbf{P1}):\: \mathop{\max}_{\mathbf{F}, \mathbf{D}} \ & \mathcal{R}=\sum_{k=1}^{K} \alpha_k \mathcal{R}_k \\
    \text{s.t.} \ & \sum_{k=1}^K\big{\Vert}\mathbf{F}\mathbf{d}_k\big{\Vert}^2\leq P, \label{wsr_total_constraints} \\
\ & \mathbf{F} \in \mathcal{F}.
    \end{align}
\end{subequations}
Here, $\mathbf{D}=\big[\mathbf{d}_1^T,\mathbf{d}_2^T,\ldots,\mathbf{d}_K^T\big]^T \in \mathbb{C}^{M \times K}$. The elements of the analog precoder are constrained by a unit modulus condition. Both digital and analog precoders are limited by the maximum transmission power constraint: $\sum_{k=1}^K||\mathbf{F}\mathbf{d}_k||^2\leq P$. This optimization task is challenging due to its non-convex objective function and complex constraints, coupled with the strong coupling between the two variables.

\subsubsection{Imperfect CSI}To optimize the hybrid precoder, accurate estimation of the mmWave channel is essential. However, in practice, due to the limited number of RF chains in the hybrid architecture, channel estimation is very challenging, which prompts the exploration of hybrid precoding in the scenarios with imperfect CSI.
To formulate this, we consider a imperfect channel estimation denoted 
\begin{equation}
\label{channel_error}
\mathbf{h}_k = \widehat{\mathbf{h}}_k + \mathbf{e}_k,
\end{equation}
with $\widehat{\mathbf{h}}_k$ and $\mathbf{e}_k$ respectively denoting the estimated channel and the corresponding error with respect to the user $k$. In this paper, we assume that the error is constrained by the spectral norm, \ie, $\sigma_\mathrm{max}(\mathbf{e}_k) \leq \epsilon_k$. Hence, when designing hybrid precoding, our aim is to make the hybrid precoder robust to estimation errors within a predefined level. This results in a maximin optimization problem:
\begin{subequations}
\label{Optimization_Problem2}
\begin{align}
(\textbf{P2}):\: \mathop{\max}_{\mathbf{F}, \mathbf{D}} \ & \min_{\sigma_\mathrm{max}(\mathbf{e}_k) \leq \epsilon_k} \sum_{k=1}^{K} \alpha_k \mathcal{R}_k \\
\text{s.t.} \ & \sum_{k=1}^K\big{\Vert}\mathbf{F}\mathbf{d}_k\big{\Vert}^2\leq P, \\
\ & \mathbf{F} \in \mathcal{F}.
\end{align}
\end{subequations}
In this formulation, we seek to maximize the minimum rate resulting from tolerable estimation error of the channel. This problem involves the joint optimization of multiple variables. Additionally, the max-min operator adds complexity to the problem.

%% file: method.tex
\vspace{0mm}

\begin{figure*}[t]
	\centering
	\begin{subfigure}[b]{0.65\textwidth}
		\includegraphics[width=\textwidth]{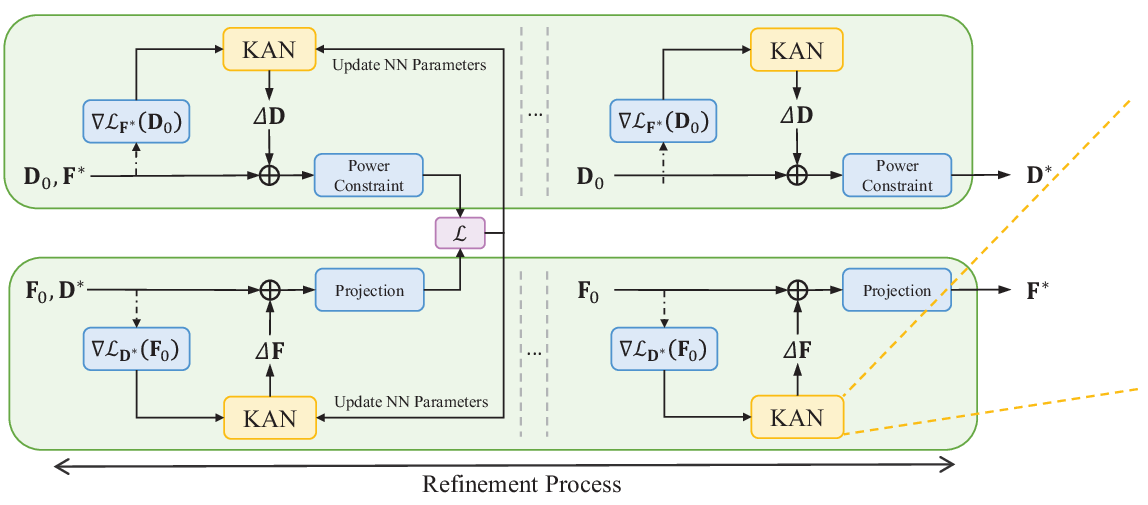}
		\caption{GGML architecture.}
		\label{fig:ggml_module}
	\end{subfigure}
        \hspace{-4mm}
	\begin{subfigure}[b]{0.3\textwidth}
		\raisebox{31pt}{\includegraphics[width=\textwidth]{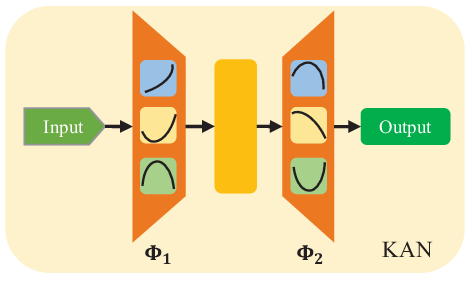}}
		\caption{Structure of a KAN.}
		\label{fig:kan}
	\end{subfigure}
	\vspace{-0mm}
	\captionsetup{font=footnotesize, name={Fig.}, labelsep=period}  
	\caption[t]{\raggedright A block diagram of the proposed meta-learning algorithm for hybrid precoding.}
	\vspace{-1mm}
	\label{fig:ggml_all}
\end{figure*}

\section{Meta Learning Based Precoding Design}\label{sec:ggml}

In this section, we describe the overall framework of GGML. We begin by introducing the built-in network KAN within GGML. Subsequently, we decompose the optimization of digital and analog precoding from a local perspective. Specifically, the digital and analog precoding matrices are refined through alternating iterations guided by gradient-based meta-learning, as illustrated in Fig. \ref{fig:ggml_all}. Ultimately, from a global perspective, the built-in network parameters are optimized with an overarching view. In the following subsections, we will detail these components.




\subsection{KAN Architecture}
KAN is a novel neural network architecture that replaces fixed activation functions with learnable activation functions at the network edge \cite{liu_kan}. It has been demonstrated in several studies that KAN achieves high accuracy in fitting complex mathematical functions \cite{liu_kan,kan_mlp}. The simplified structure of the network is shown in Fig. \ref{fig:kan}. This network is inspired by the Kolmogorov-Arnold Representation theorem, which states that any multivariate continuous function on a bounded domain can be expressed as a finite composition of continuous single-variable functions and binary operations of addition. Specifically, it can be represented as
\begin{equation}
    \label{Kolmogorov-Arnold}
    f(x_1,\cdots,x_n)=\sum_{q=1}^{2n+1}\Phi_q\Bigg(\sum_{p=1}^n\phi_{q,p}\big(x_p\big)\Bigg),
\end{equation}
where $f:[0,1]^{n}\to\mathbb{R}$, $\phi_{q,p}:[0,1]\to\mathbb{R}$ and $\Phi_q:\mathbb{R}\to\mathbb{R}$. 
Similar to an MLP, A $K$-layer KAN can be described as a nesting of multiple KAN layers:
\begin{equation}
    \label{KANforward}
    \mathrm{KAN}\big(\mathbf{Z}\big)=\big(\mathbf{\Phi}_{K-1} \circ \mathbf{\Phi}_{K-2} \circ \ldots \circ \mathbf{\Phi}_{0}\big)\mathbf{Z},
\end{equation}
where $\mathbf{\Phi}_{i}$ represents the $i$-th layer of KAN, consisting of $n_{in}\times n_{out}$ learnable activation functions, namely, 
\begin{equation}\label{activation_function}
    \mathbf{\Phi}=\{\phi_{i,q,p}\} ,\  p=1,2,\cdots,n_{\mathrm{in}} ,\  q=1,2\cdots,n_{\mathrm{out}} .
\end{equation}

To better fit functions, KAN uses B-splines as part of the activation function. The flexibility of the splines allows them to adaptively model complex relationships in the data by adjusting their shape, thereby minimizing approximation error. This enhances the network's ability to learn subtle patterns from high-dimensional datasets. It can be represented as
\begin{equation}\label{B-spline}
    \mathrm{spline}\big(x\big)=\sum_{i}c_iB_i\big(x\big),
\end{equation}
Here, $\mathrm{spline}\big(x\big)$ represents the spline function. $c_i$ are learnable parameters, and $B_i\big(x\big)$ are the B-spline basis functions defined on a grid. The grid points define the intervals where each basis function $B$ is active and significantly influences the shape and smoothness. Given the described spline function, the activation function of the KAN can be written as
\begin{equation}\label{activation_function2}
    \phi(x)=w_bb(x)+w_s\mathrm{spline}(x),
\end{equation}
where $b(x)=\mathrm{silu}(x)=x/(1+e^{-x})$. 
It can be seen that KAN retains more feature information through residual connections.

\subsection{GGML Overall Framework}
\subsubsection{Gradient Based Optimization}Many DL-based precoding strategies typically take the channel matrix as the input to the neural network and directly use the output as the optimal precoding matrix. However, such strategies often lead to a decline in overall performance: First, it lacks interpretability, as it cannot explain the black-box operations of the NN. Second, traditional neural network architectures struggle to capture optimal information from high-dimensional channel matrices and are highly sensitive to channel estimation errors. To address this issue, we retain gradient optimization, using the gradient of the optimization objective as the input to the neural network and updating based on the gradient information flow from the output. This approach combines the strengths of neural networks in feature extraction with the capabilities of meta-learning in integrating information flows, potentially enhancing optimization performance. Moreover, the introduction of gradient information also improves the interpretability of the model.
\subsubsection{Local Perspective Based Optimization}The updates for the subproblems rely on a local perspective, drawing on the idea of alternating optimization. By fixing one precoding matrix parameter in each iteration and optimizing the other, the coupling between variables is progressively reduced. We first initialize the analog and digital precoding matrices as $\mathbf{F}_0$, $\mathbf{D}_0$ respectively. The corresponding neural networks for the subproblems are the digital precoding network (DPN) and the analog precoding network (APN).
\begin{itemize}
    \item \emph{Digital Precoding Update:} For a clear illustration, we first present the properties of the full power constraint through the following proposition.
    \begin{proposition}[Full Power Property]
    \label{proposition_optimal_with_equality}
    When the analog precoding matrix $\mathbf{F}$ is fixed, the optimal solution $\{\mathbf{D}^*\}$ to problem \eqref{Optimization_Problem1} satisfy the sum power constraint \eqref{wsr_total_constraints} with equality.
    \end{proposition}
    \par
    \begin{proof}
    Analogous to the proof process in \cite{zfh_meta} and \cite{rethinking_wmmse}, by leveraging the Karush-Kuhn-Tucker (KKT) condition, it can be deduced that the Lagrange multipliers must be positive. Consequently, through the application of the complementary slackness condition, it follows that the optimal solution must satisfy the equality power constraint.
    \end{proof}
    \par
    In the process of updating the digital precoder, we first calculate the loss function and its gradient with respect to $\mathbf{D}_0$ as $\nabla\mathcal{L}_{\mathbf{F}^*}\big(\mathbf{D}_0\big) \in \mathbb{C}^{M \times K}$, where $\mathbf{F}^*$ represents the initialized or updated precoding matrix. Subsequently, we consider the $\nabla\mathcal{L}_{\mathbf{F}^*}\big(\mathbf{D}_0\big)$ as a one-dimensional vector of size $MK$ and feed them into the KAN, where the network parameters are kept frozen. Both the input and output are decomposed into real and imaginary components, so the input and output layers of the KAN consist of $2MK$ neurons each. The update criterion can be expressed as follows: 
    \begin{equation}
        \label{update_ruler}
        {\mathbf{D}} = {\mathbf{D}_0} + \mathrm{DPN}\Big(\nabla\mathcal{L}_{\mathbf{F}^*}\big(\mathbf{D}_0\big)\Big), 
    \end{equation}
    where DPN serves as an optimizer for variable updates, with its learnable parameters denoted by $\theta_{\mathbf{D}}$.
    \par
    By using Proposition \ref{proposition_optimal_with_equality}, we can constrain the optimized digital precoder to the maximum achievable transmit power, denoted as
    \begin{equation}
        \label{constrain_max_power}
        {\mathbf{D}^*} = \sqrt{\frac{P}{\big{\Vert}\mathbf{F}\mathbf{D}\big{\Vert}^2}}\mathbf{D}.
    \end{equation}
    \par
    \item \emph{Analog Precoding Update:} Similar to the update of the digital precoder, we compute the loss function and its gradient with respect to $\mathbf{F}_0$ as $\nabla\mathcal{L}_{\mathbf{D}^*}\big(\mathbf{F}_0\big) \in \mathbb{C}^{N \times M}$ and feed it into the APN. However, the constant modulus constraint of the analog precoder results in a fundamentally different optimization behavior for the target variable, posing a significant challenge. Therefore, we consider employing projected gradient ascent (PGA) algorithm to address the constant modulus constraint. In this iterative method, each iteration first optimizes $\mathbf{F}$ based on GGML, and then projects the optimized precoding matrix to ensure that the constant modulus constraint is not violated.. This design ensures the convergence and stability of $\mathbf{F}$ during the inner iteration process, which can be expressed as 
    \begin{equation}
        {\mathbf{F}} = \mathcal{P}_{\mathcal{F}}\left({\mathbf{F}_0} + \mathrm{APN}\Big(\nabla\mathcal{L}_{\mathbf{D}^*}\big(\mathbf{F}_0\big)\Big)\right), \label{update_ruler_ana}
    \end{equation}
    where $\mathcal{P}_{\mathcal{F}}$ is the projection operator onto the set ${\mathcal{F}}$. 
    \par
    In \eqref{update_ruler_ana}, the projection operator $\mathcal{P}_{\mathcal{F}}$ depends on the feasible analog mappings. For unconstrained architectures, $\mathcal{P}_{\mathcal{F}}(\mathbf{F})=\mathbf{F}$. For fully connected architecture, the projection is given as 
    \begin{equation}
        \label{project}
        \mathcal{P}_{\mathcal{F}}\big(\mathbf{F}\big)=\Tilde{\mathbf{F}},\quad [\Tilde{\mathbf{F}}]_{i,j}=\frac{[\mathbf{F}]_{i,j}}{\big{\vert}[\mathbf{F}]_{i,j}\big{\vert}}.
    \end{equation}
\end{itemize}
\begin{algorithm}[tb]
  \SetKwInOut{Input}{Input}\SetKwInOut{Output}{Output}
  \Input{$\mathbf{h}_{ k}$, $\alpha_k$, for $k = 1,2,\ldots, K$}
  Randomly initialize $\mathbf{F}$\label{alg1_init1}\;
  Initialize $\mathbf{D}$ with ZF beamforming\label{alg1_init2}\;
  Set $ i = 1$\;
  \While{\textbf{$i \leq L$} \label{alg1_iter0}}{
    Update $\mathbf{F}$ according to (\ref{update_ruler_ana})\;
    Update $\mathbf{D}$ according to (\ref{update_ruler}) and scale $\mathbf{D}$ according to the maximum transmit power constraint\;
    Calculate the $\mathcal{L}$ as \eqref{loss}\;
    Update $\mathbf{\theta_D}$ and $\mathbf{\theta_F}$ according to \eqref{update_D_NN_parameter} and \eqref{update_F_NN_parameter}\;
    $i = i+1$\;
  }\label{alg1_iter4}
  \Output{$\mathbf{F}$, $\mathbf{d}_{k}$, for $k = 1,2,\ldots, K$}
  \caption{The Overall Framework of the Proposed GGML for Hybrid Precoding with Perfect CSI}
  \label{alg_WMMSE_PINet_Perfect}
\end{algorithm}
\par
\subsubsection{Global Perspective Based Optimization}The meta-learning network parameters are updated from a global perspective. We adopt a loss function with a penalty term as the global loss of the system, which can be denoted as
\begin{equation}
    \label{loss}
    \mathcal{L}=-\mathcal{R}+\beta \cdot \mathrm{Var}\big(\mathcal{R}\big),
\end{equation}
where $\mathcal{L},\beta$ denote the loss function, penalty rate, respectively. Then backward propagation is conducted with Adam optimizer, as depicted below
\begin{align}
    \mathbf{\theta_D}^* & =\mathbf{\theta_D} + \alpha_{\mathbf{D}} \cdot \mathrm{Adam}(\nabla_{\mathbf{\theta_D}}{\mathcal{L}}, \mathbf{\theta_{D}}), \label{update_D_NN_parameter} \\
    \mathbf{\theta_F}^* & =\mathbf{\theta_F} + \alpha_{\mathbf{F}} \cdot \mathrm{Adam}(\nabla_{\mathbf{\theta_F}}{\mathcal{L}}, \mathbf{\theta_{F}}). \label{update_F_NN_parameter}
\end{align}
where $\mathbf{\theta_D}$ and $\mathbf{\theta_F}$ are the learning rates of DPN and APN, respectively.
\par
Through this global perspective optimization, GGML not only focuses on the local adjustments of each network layer during the update process but also emphasizes how minimizing the global loss drives the optimization of the entire hybrid precoding architecture. Specifically, GGML integrates the synergy and mutual influence between the digital and analog precoders, establishing new update clues for hybrid precoding optimization. Compared to traditional alternating optimization methods, GGML leverages a meta-learning mechanism, enabling it to gradually learn how to more efficiently adjust precoding parameters over multiple iterations, allowing the system to rapidly adapt to dynamic environments while maintaining high performance. This global perspective-based optimization effectively overcomes performance bottlenecks caused by local optima in traditional methods, ensuring near-global optimal hybrid precoding performance in large-scale MIMO systems. The GGML based hybrid precoding algorithm is  outlined in Algorithm \ref{alg_WMMSE_PINet_Perfect}.
\par
Meanwhile, GGML’s global optimization mechanism reduces computational complexity and training time, as it avoids repeatedly solving local problems in each iteration. Instead, it optimizes the entire system’s performance through a few key steps. This efficient learning mechanism provides new solutions for future wireless communication systems, especially for hybrid precoding problems in mmWave communications.
\subsection{Complexity Analysis}In this subsection, we compare the complexity of the proposed algorithm with other hybrid precoding algorithms, including the MO-alternating minimization (MO) algorithm \cite{yxhhybrid}, the Element-alternating maximization (Element-AltMax) algorithm \cite{yuweihybrid}, the CNN algorithm \cite{jhhybrid} and the majorization minimization-alternating minimization algorithm (MM) \cite{hscMM-Altmin}. The complexities of the algorithms are listed in Table \ref{tab:complexity}, where $\{L,T\}$ represent the number of iterations in the algorithms. In the proposed GGML algorithm, the computational complexity mainly arises from the update of the digital precoder and the analog precoder. The update of the digital precoder is divided into three steps. The first step includes the calculation of the loss function and gradient, with a complexity of $\mathcal{O}(NMK^2)$; the second step includes the forward calculation of the network, with the complexity of $\mathcal{O}(NM)$; the third steps includes the power normalization in \eqref{constrain_max_power}, with a complexity of $\mathcal{O}(NMK)$. In conclusion, the computational complexity of the update of the digital precode is $\mathcal{O}(NMK^2)$. Similarly, the complexity of the analog precoding update is $\mathcal{O}(NMK^2)$. Given that these operations are performed $L$ times per sample, the overall complexity of the GGML is $\mathcal{O}(LNMK^2)$. 
\par
In mmWave MIMO systems, the number of antennas is usually large for adequate array gain, so the term that mainly contributes to the computational complexity is the one which contains $N$. From Table \ref{tab:complexity}, it can be observed that except for the proposed GGML algorithm and CNN, other algorithms all include quadratic or cubic terms of $N$.

%% file: imcsi.tex
\vspace{0mm}
\begin{table}[t]
\centering
\caption{\scriptsize COMPLEXITY ANALYSIS}\vspace{0mm}
\begin{tabular}{c c} 
\toprule
Algorithm & Computational Complexity \\
\midrule
MO & $\mathcal{O}(LN^2MK)$\\
Element-AltMax & $\mathcal{O}(LN^3MK+M^2K+TK)$\\
CNN & $\mathcal{O}(MK)$\\
MM & $\mathcal{O}(L(TKN^3M^3+NM^2))$\\
GGML & $\mathcal{O}(LNMK^2)$\\
\bottomrule
\end{tabular}\vspace{-3mm}
\label{tab:complexity}
\end{table}
\section{Extension to Imperfect CSI Scenarios}\label{sec:imcsi}
This section introduces the extension of the proposed meta-learning-based precoding algorithm, aimed at addressing the worst-case weighted sum rate maximization problem, denoted as $\textbf{P2}$. Motivated by the Worst-Case framework outlined in \cite{worst_case}, we reformulate the worst-case weighted sum rate maximization problem of downlink hybrid precoding. Meanwhile, the GGML approach is adopted to solve the worst-case performance optimization problem.

\begin{figure*}[t]
\centering
\includegraphics[width=0.9\linewidth]{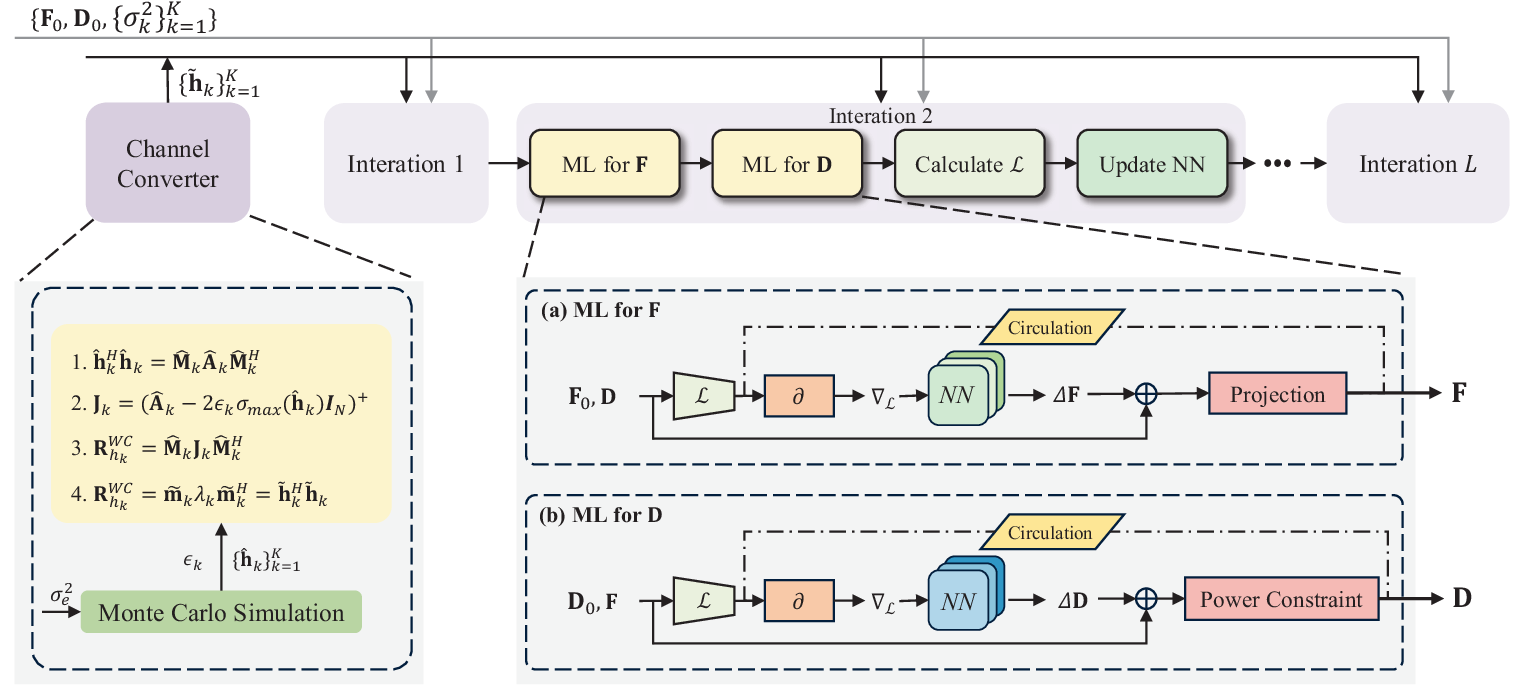}
\captionsetup{font=footnotesize, name={Fig.}, labelsep=period}  
\caption{\raggedright Schematic diagram of the proposed GGML-ImCSI for robust downlink hybrid precoding.}
\label{fig:ggml_robust}
\end{figure*}

\par
\subsection{Problem Transformation}The optimization problem $\textbf{P2}$ is challenging to be solved because the inner minimization is not tractable. We first introduces the equivalent mean square error of the received signal $s_k$, which can be written as
    
\begin{equation}
\begin{aligned}
\label{MSE}
    e_k &= 1 - u_k^\ast \mathbf{h}_k \mathbf{F} \mathbf{d}_k - \mathbf{d}_k^H \mathbf{F}^H \mathbf{h}_k^H u_k \\
    &\quad + \sum_{m=1}^{K} u_k^\ast \mathbf{h}_k \mathbf{F} \mathbf{d}_k \mathbf{d}_k^H \mathbf{F}^H \mathbf{h}_k^H u_k + u_k^\ast u_k,
\end{aligned}
\end{equation}
where $u_k$ is the receive gain of user $k$.
\par
Fixing $\mathbf{F}$ and $\mathbf{D}$, the solution of the optimal $u_k$ is given by 
\begin{equation}
    \label{MMSE rec}
    u_k^{mmse}={\Bigg(\mathbf{h}_k\mathbf{F}\sum_{m=1}^{K}{\mathbf{d}_m\mathbf{d}_m^H}\mathbf{F}^H\mathbf{h}_k^H+\sigma_k^2\Bigg)}^{-1}\mathbf{h}_k\mathbf{F}\mathbf{d}_k.
\end{equation}
\par
By substituting $u_k^{mmse}$ back into \eqref{MSE}, the minimum mean square error (MMSE) becomes 
\begin{equation}
    \label{MMSE}
    e_k^{mmse}=1-\mathbf{d}_k^H \mathbf{F}^H \mathbf{h}_k^H{b_k}^{-1}\mathbf{h}_k\mathbf{F}\mathbf{d}_k,
\end{equation}
where
\begin{equation}
    {b_k}={\Bigg(\mathbf{h}_k\mathbf{F}\sum_{m=1}^{K}{\mathbf{d}_m\mathbf{d}_m^H}\mathbf{F}^H\mathbf{h}_k^H+\sigma_k^2\Bigg)}
\end{equation}
\par
Thus, according to [21], the equivalent mean square error is written as 
\begin{equation}
    \label{eq-MMSE}
    e_k^{mmse}={\Big(1+\mathbf{d}_k^H \mathbf{F}^H\mathbf{R}_{{h}_k}{\big(\mathbf{T}_k\mathbf{R}_{{h}_k}+\sigma_k^2\mathbf{I}_N\big)}^{-1}\mathbf{F}\mathbf{d}_k\Big)}^{-1},
\end{equation}
where $\mathbf{R}_{h_k}=\mathbf{h}_k^H\mathbf{h}_k$ and $\mathbf{T}_k \triangleq \sum_{m=1,m\neq k}^{K}{\mathbf{F}\mathbf{d}_m\mathbf{d}_m^H\mathbf{F}^H}$. Thus, the achievable rate of the user $k$ is expressed as
\begin{equation}
    \label{eqrate}
    \mathcal{R}_k=-\log\left(e_k^{mmse}\right).
\end{equation}
\par
Subsequently, we reformulate the maximin problem $\textbf{P2}$. First of all, we substitute \eqref{channel_error} into $\mathbf{R}_{h_k}$ and assume $\sigma_{max}({\widehat{\mathbf{h}}}_k)\geq\epsilon_k$. By neglecting the second order term of the error, $\mathbf{R}_{h_k}$ can be written as 
\begin{equation}
    \label{eq-Rhk}
    \mathbf{R}_{h_k}={({\widehat{\mathbf{h}}}_k+\mathbf{e}_k)}^H({\widehat{\mathbf{h}}}_k+\mathbf{e}_k)\approx{\widehat{\mathbf{R}}}_{h_k}+\mathbf{R}_{{he}_k},
\end{equation}
Here, ${\widehat{\mathbf{R}}}_{h_k}={\widehat{\mathbf{h}}}_k^H{\widehat{\mathbf{h}}}_k$ and $\mathbf{R}_{{he}_k}=\mathbf{e}_k^H{\widehat{\mathbf{h}}}_k+{\widehat{\mathbf{h}}}_k^H\mathbf{e}_k$.

\par
To reformulate the worst-case weighted sum rate maximization problem, we have the following essential facts summarized in Lemma~\ref{lemma_with_Rh}.
\begin{lemma}[adapted from Lemma 7.1 in \cite{palomar2003unified}]
\label{lemma_with_Rh}
For matrices $\mathbf{X}\in\mathbb{C}^{n\times m}$ and $\mathbf{Y}\in\mathbb{C}^{n\times m}$, the following inequality holds: 
\begin{equation}
    \label{XY}
    \mathbf{X}\mathbf{Y}^H + \mathbf{Y}\mathbf{X}^H \leq 2\sigma_{\max}(\mathbf{X})\sigma_{\max}(\mathbf{Y})\mathbf{I}_n
\end{equation}
\end{lemma}
Then by applying Lemma~\ref{lemma_with_Rh} to $\mathbf{R}_{h_k}$, we obtain that $\mathbf{R}_{h_k}$ is bounded as 
\begin{equation}
    \label{Rh_bound}
    \mathbf{R}_{h_k}\succeq\mathbf{R}_{h_k}^{WC} \triangleq \mathbf{\widehat{M}}_{k}\mathbf{J}_{k}\mathbf{\widehat{M}}_{k}^{H},
\end{equation}
where $\mathbf{J}_{k}=\Big(\widehat{\mathbf{A}}_{k}-2\epsilon_{k}\sigma_{max}\big(\widehat{\mathbf{h}}_{k}\big)\mathbf{I}_{N}\Big)^{+}$, $\mathbf{\widehat{M}}_{k}$ and $\widehat{\mathbf{A}}_{k}$ are computed by the eigenvalue decomposition of ${\widehat{\mathbf{R}}}_{h_k}$. Otherwise, since $\rank\left({\widehat{\mathbf{h}}}_k\right)=\rank\left(\mathbf{J}_k\right)=1$, we can reformulate $\mathbf{R}_{h_k}^{WC}$ as 
\begin{equation}
    \label{refor-rhwc}
    \mathbf{R}_{h_k}^{WC}={\widetilde{\mathbf{m}}}_k\lambda_k{\widetilde{\mathbf{m}}}_k^H={\widetilde{\mathbf{h}}}_k^H{\widetilde{\mathbf{h}}}_k,
\end{equation}
where ${\widetilde{\mathbf{m}}}_k$ represents the rightmost vector of $\mathbf{\widehat{M}}_{k}$, $\mathbf{\widehat{M}}_{k}$ is the unique non-zero eigenvalue of $\mathbf{J}_{k}$ and ${\widetilde{\mathbf{h}}}_k=\sqrt{\lambda_k}{\widetilde{\mathbf{m}}}_k^H$.
\par
The relationship between $e_k^{mmse}$ and $\mathbf{R}_{h_k}$ is established in the following proposition. 
\par
\begin{proposition}\label{mmse_gradient}
For any combination $\mathbf{T}_k$ of hybrid precoder, $e_k^{mmse}$ can achieve an upper bound at the point $\mathbf{R}_{h_k}^{WC}$.
\end{proposition}
\par
\begin{proof}
    The proof is provided in Appendix \ref{appendix1}.
\end{proof}
By applying Proposition \ref{mmse_gradient}, we have
\begin{equation}
    \label{mmse_wc}
    e_k^{WC}={(1+\mathbf{d}_k^H\mathbf{F}^H\mathbf{R}_{h_k}^{WC}{(\mathbf{T}_k\mathbf{R}_{h_k}^{WC}+\sigma_k^2\mathbf{I}_N)}^{-1}\mathbf{F}\mathbf{d}_k)}^{-1}.
\end{equation}
\par
Subsequently, we can transform the maximin optimization into the problem of minimizing the GP of Worst-Case MMSE
\begin{subequations}
\label{Optimization_Problem3}
\begin{align}
(\textbf{P3}):\: \mathop{\min}_{\mathbf{F}, \mathbf{D}} \ & \mathcal{E}=\prod_{k=1}^{K}{(e_{k}^{WC})}^{\alpha_k} \\
\text{s.t.} \ & \sum_{k=1}^K\big{\Vert}\mathbf{F}\mathbf{d}_k\big{\Vert}^2\leq P, \\
\ & \mathbf{F} \in \mathcal{F}.
\end{align}
\end{subequations}
\par
Interestingly, the optimization problem \textbf{P3} can be regarded as a MIMO hybrid precoding problem based on perfect CSI by replacing $\{\mathbf{h}_k~\text{with}~{\widetilde{\mathbf{h}}}_k\}_{k=1}^{K}$.

\begin{algorithm}[t]
  \SetKwInOut{Input}{Input}\SetKwInOut{Output}{Output}
  \Input{$\mathbf{h}_{ k}$, $\alpha_k$, $\epsilon_{k}$, for $k = 1,2,\ldots, K$}
  Calculate $\mathbf{R}_{h_k}^{WC}$ according to \eqref{Rh_bound} by the eigenvalue decomposition\;
  Calculate ${\widetilde{\mathbf{h}}}_k$ according to \eqref{refor-rhwc}\;
  Randomly initialize $\mathbf{F}$\label{alg2_init1}\;
  Initialize $\mathbf{D}$ with ZF beamforming\label{alg2_init2}\;
  Set $ i = 1$\;
  \While{\textbf{$i \leq L$} \label{alg2_iter0}}{
    Update $\mathbf{F}$ according to (\ref{update_ruler_ana})\;
    Update $\mathbf{D}$ according to (\ref{update_ruler}) and scale $\mathbf{D}$ according to the maximum transmit power constraint\;
    Calculate the $\mathcal{L}$ as \eqref{loss-robust}\;
    Update $\mathbf{\theta_D}$ and $\mathbf{\theta_F}$ according to \eqref{update_D_NN_parameter} and \eqref{update_F_NN_parameter}\;
    $i = i+1$\;
  }\label{alg2_iter4}
  \Output{$\mathbf{F}$, $\mathbf{d}_{k}$, for $k = 1,2,\ldots, K$}
  \caption{GGML-ImCSI for Hybrid Precoding with Imperfect CSI}

  \label{alg_WMMSE_PINet_ImPerfect}
\end{algorithm}
\vspace{-2mm}
\subsection{Meta Learning Based Robust Precoding Design}
In this subsection, we propose an extension of GGML to handle imperfect CSI scenarios, referred to as GGML-ImCSI, with its schematic structure shown in the Fig. \ref{fig:ggml_robust} and presented in Algorithm \ref{alg_WMMSE_PINet_ImPerfect}. The first module is the channel converter node, which implements the steps for transforming \textbf{P2} into \textbf{P3}. Therefore, it can compute the approximate channel ${\widetilde{\mathbf{h}}}_k$ using the inputs ${{\widehat{\mathbf{h}}}_k}$ and $\epsilon_{k}$. For worst-case robust precoding optimization, we must determine the maximum spectral norm of the channel error vector in advance. We adopt the method from \cite{worst_case}, where $10^5$ Monte Carlo error realizations are generated, and the spectral norm is determined based on a 
$5\%$ outage probability, \ie, $\Pr\{\sigma_{max}(\mathbf{e}_k)\geq\epsilon\}\leq 5\%$.
\par
The logic of GGML-ImCSI is similar to GGML, where the precoding matrix is optimized from a local perspective, and the network's built-in parameters are optimized from a global perspective. However, in GGML-ImCSI, the selected loss function differs due to the change in the objective function. In each iteration step, the product of the MMSE is directly used as the loss function, defined as 
\begin{equation}
    \label{loss-robust}
    \mathcal{L}=\mathcal{E}+\beta\cdot\mathrm{Var}(\mathcal{E}).
\end{equation}

%% file: simulation.tex
\vspace{1mm}

\section{Simulation Results}\label{sec:simulation}
In this section, we evaluate the performance of the proposed hybrid precoding algorithm through numerical simulations. We begin by introducing the specific details concerning the simulation parameters and the comparison algorithms. Subsequently, we present the performance results of the algorithm across a range of system configurations. Finally, we examine the performance of algorithm in practical scenarios with imperfect CSI.
\par
\subsection{Simulation Details}
The channel models used in our simulation are as defined in \eqref{channel}. The complex gain of each ray within a cluster follows a standard complex Gaussian distribution, while the departure angles are Laplace-distributed within the range of \(0\) to \(2\pi\), capturing the variations in propagation paths among rays within each cluster. The imperfect channel model is given by \eqref{channel_error}, where the variable \(\mathbf{e}_k\) denotes the estimation error, with elements that follow a Gaussian distribution with zero mean, i.e., \(e \sim \mathcal{C}N\left(0, \sigma_e^2\right)\). The variance \( \sigma_e^2 \) (representing the error power) is defined as \( \sigma_e^2 \triangleq \delta |h|^2 \), where \(\delta\) quantifies the ratio between the error power \( \sigma_e^2 \) and the channel gain \( |h|^2 \), thereby characterizing the degree of CSI error\footnote{A statistical CSI error model is adopted here, appearing within the objective function as a max-min operation rather than as a constraint.}.
\par
To streamline the simulation, we assume equal noise variances for the received signals of all users, \ie, \(\sigma_1^2 = \sigma_2^2 = \cdots = \sigma_K^2=\sigma^2\). Meanwhile, the maximum transmit power is set to 1, and SNR in dB is defined by the formula \(10\log_{10}(1/\sigma^2)\). To ensure fairness in resource allocation, each user is assigned the same priority, specifically \(\alpha_1 = \alpha_2 = \cdots = \alpha_K = 1 / K\). Consequently, average spectral efficiency is used as the primary performance metric. In the simulation, our proposed algorithm is compared with several benchmarks, with results averaged over \(10^2\) independent channel realizations. All simulations are executed on a PC running Windows 10 with an Intel Core i7-14700 CPU @ 3.40 GHz. The system parameters \(L\), \(\alpha_{\mathbf{D}}\), \(\alpha_{\mathbf{F}}\), \(N_{\mathrm{c}}\), and \(N_{\mathrm{ray}}\) are set to \(500\), \(1\times 10^{-3}\), \(1.5\times 10^{-3}\), \(3\), and \(30\), respectively. The deep learning-based model is implemented in PyTorch, using the Adam optimizer for network parameter updates.
\par
\subsection{Convergence Analysis}
\label{subsec:convergence}
\begin{figure}[t]\vspace{-0mm}  
	\begin{center}		\centerline{\includegraphics[width=0.42\textwidth]{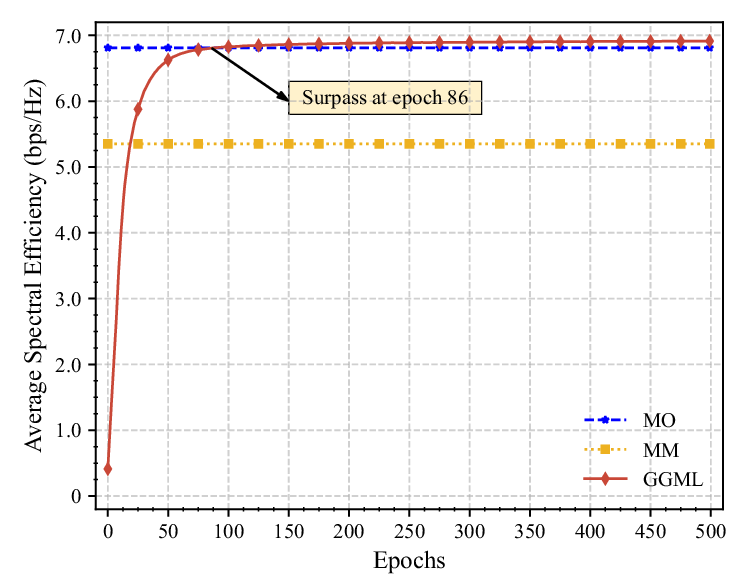}}  \vspace{-0mm}
	    \captionsetup{font=footnotesize, name={Fig.}, labelsep=period}  
        \caption[t]{\raggedright Average spectral efficiency versus the iterations.}
        
		\label{fig:convergence_result} \vspace{-4mm}
	\end{center}
\end{figure}

To show the convergency of the proposed algorithm, we plot the average spectral efficiency against the number of epochs in Fig. \ref{fig:convergence_result} by running algorithms with \(10^2\) random initializations. The simulation setups are as follows: \(N=64\), \(M=K=4\), and the SNR is fixed at \(\text{SNR}=10\text{dB}\). The results in Fig. \ref{fig:convergence_result} illustrate that the proposed hybrid precoding framework can converge within 200 epochs. To be more specific, the performance of {\it GGML} steadily improves with increasing iterations, surpassing traditional methods at epoch 86. The essential reason is that, global optimization strategy in {\it GGML} could effectively reduces the complexity of solving local sub-problems, facilitating convergence to the optimal solution.
\par
Moreover, after 500 epochs, the {\it GGML} algorithm outperforms {\it MO} and {\it MM} by 1.4\% and 42.4\%, respectively. The superior performance of {\it GGML} is largely attributed to its efficient combination of local and global optimization techniques.

\begin{figure}[t]\vspace{-0mm}
	\begin{center}		\centerline{\includegraphics[width=0.42\textwidth]{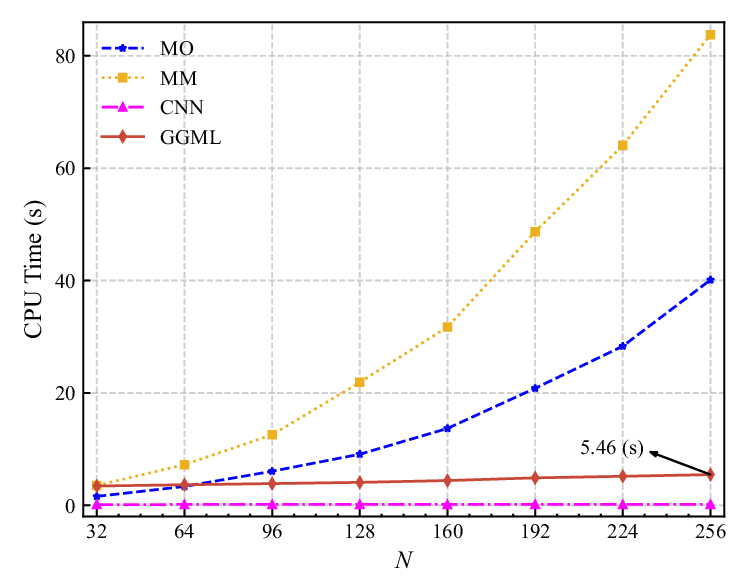}}  \vspace{-0mm}
	    \captionsetup{font=footnotesize, name={Fig.}, labelsep=period}  
        \caption[t]{\raggedright Average CPU time versus the BS antenna.}
		\label{fig:run_time} \vspace{-4mm}
	\end{center}
\end{figure}
\subsection{Complexity Comparison}
To compare the computational complexity of different algorithms more intuitively, we evaluate the average CPU time under varying numbers of BS antennas in this subsection. As shown in Fig. \ref{fig:run_time}, the CPU time consumption of different algorithms exhibits significant differences as the number of antennas, \( N \), increases. Specifically, traditional algorithms demonstrate a pronounced nonlinear increase in CPU time with an increasing number of antennas. In contrast, neural network-based algorithms are less sensitive to variations in antenna numbers, achieving remarkably high computational efficiency. Although {\it CNN}-based algorithms can achieve millisecond-level computation times, they require extensive offline training and still underperform compared to {\it GGML}, which will be discussed in detail in the following subsection. Notably, when \( N = 256 \), the {\it GGML} algorithm reduces the runtime by approximately 80\% compared to {\it MO}, indicating that our algorithm is more than 8 times faster than traditional state-of-the-art methods, which is consistent with the previous complexity analysis.
\par
\subsection{Performance Analysis}\label{subsec:performance}

\begin{figure*}[ht!]
    \centering
    \begin{subfigure}[b]{0.42\textwidth}
        \includegraphics[width=\textwidth, clip]{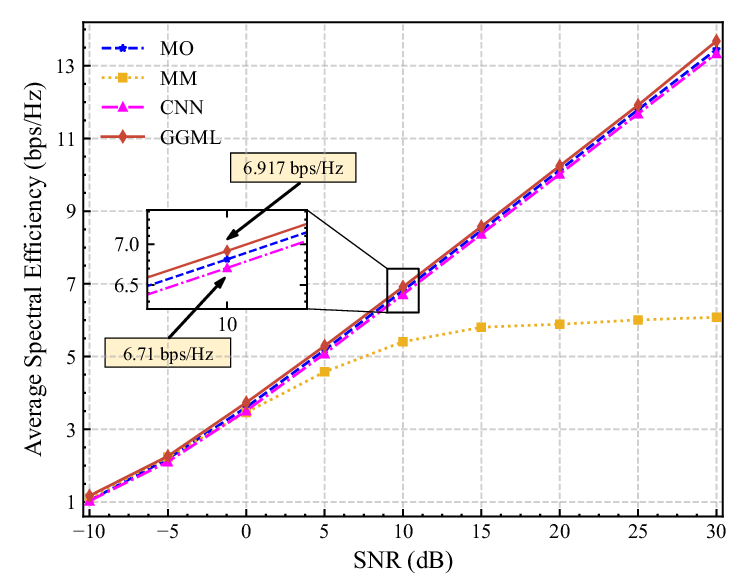}
        \caption{}
        \label{fig_robustness_a}
    \end{subfigure}
    \hspace{8mm} 
    \begin{subfigure}[b]{0.42\textwidth}
        \includegraphics[width=\textwidth, clip]{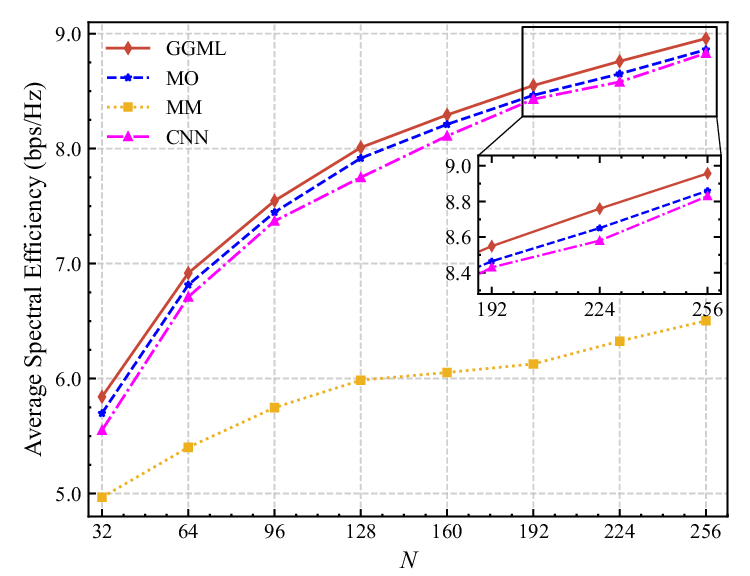}
        \caption{}
        \label{fig_robustness_b}
    \end{subfigure}
    \begin{subfigure}[b]{0.42\textwidth}
        \includegraphics[width=\textwidth, clip]{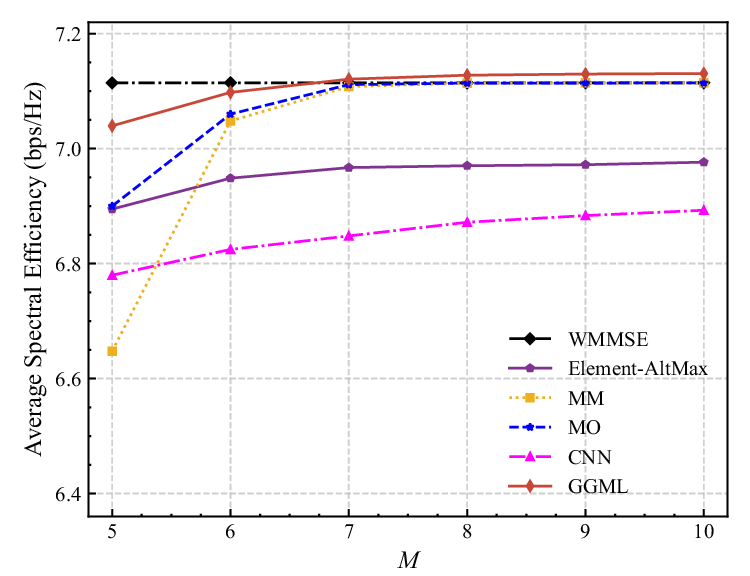}
        \caption{}
        \label{fig_robustness_c}
    \end{subfigure}
    \hspace{8mm}
    \begin{subfigure}[b]{0.42\textwidth}
        \includegraphics[width=\textwidth, clip]{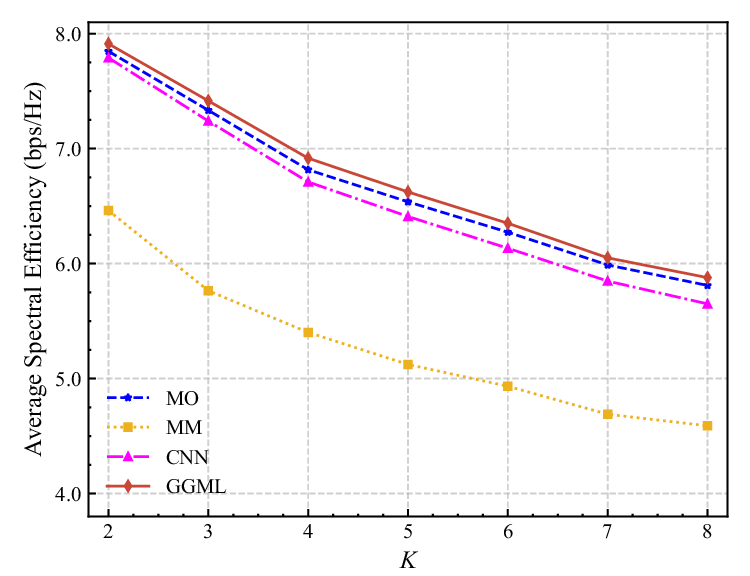}
        \caption{}
        \label{fig_robustness_d}
    \end{subfigure}
    \captionsetup{font=footnotesize, name={Fig.}, labelsep=period} 
    \caption{The performance with different system parameters, (a) SNR, (b) number of BS antennas, $N$, (c) number of RF chains, $M$, and (d) number of users, $K$.}
    \label{fig_Performance}
    \vspace{-4mm}
\end{figure*}
In order to evaluate the performance of the proposed scheme in the mmWave downlink MU-MISO system, we conducted simulations across various scenarios encompassing distinct SNR, numbers of BS antennas, numbers of RF chains and numbers of users $K$. As shown in Fig. \ref{fig_Performance}, the proposed {\it GGML} algorithm consistently outperforms other algorithms across all scenarios by leveraging the inherent strengths of gradient descent and meta-learning.
\par
Fig. \ref{fig_robustness_a} present the average spectral efficiency against the SNR, with \(N\), \(M\), and \(K\) fixed at 64, 4, and 4, respectively. From this figure we can observe that, with the increase of the SNR, the average spectral efficiency rises rapidly in all cases. Particularly, {\it GGML} outperforms {\it MM} significantly and surpasses {\it MO} and {\it CNN} slightly. This phenomenon can be attributed to the aptitude of {\it GGML} to escape local optima during joint optimization, a task that is arguably accomplished more effectively than traditional alternating optimization methods and data-driven deep learning approaches.
\par
In Fig. \ref{fig_robustness_b}, the average spectral efficiency is plotted against the number of BS antennas using the same setups as above. We can observe that, all schemes benefit from an increase in the number of antennas. More importantly, we find that, with the increasing of BS antennas, the effectiveness of NNs becomes more pronounced. For instance, when $N=128$, the spectral efficiency of the {\it CNN} reaches approximately 98\% of the optimal solution, while at $N=256$, this efficiency rises to 99.6\%. The observation implies that, NNs can efficiently extract features from high-dimensional optimization spaces. Furthermore, due to the effective integration of meta-learning and gradient flow, {\it GGML} demonstrates a significant performance improvement over {\it CNN}.
\par
To evaluate the performance of the proposed hybrid precoding scheme with more RF chains, in this part, we analyze the system average spectral efficiency with different numbers o RF chains in Fig. \ref{fig_robustness_c}. From this figure, we can observe that, the average spectral efficiency increases with the stronger signal processing capabilities of the digital precoder. When \(M > 6\), {\it GGML} even outperforms {\it WMMSE} precoding (simulated as a fully digital system with the same number of antennas). This outcome suggests that an optimization approach with both local and global perspectives can significantly enhance precoding performance, allowing it to exceed fully digital precoders with only a small number of RF chains. Thus, we can conclude that the proposed {\it GGML} scheme can serve as an effective solution for mmWave hybrid precoding, particularly suitable for use in Massive MIMO systems with a large number of antennas and a limited number of RF chains.
\par
Then, using the same setups in Subsection \ref{subsec:convergence}, we plot the average spectral efficiency versus the number of users \(K\) in Fig. \ref{fig_robustness_d} where the number of RF chains is kept consistent with the number of users. We can observe that, the performance loss grows with the increasing of \(K\) and the proposed {\it GGML} consistently surpasses other algorithms across different user counts. This phenomenon can be explained by two factors as follows. First of all, the {\it GGML} incorporates SE variance as a constraint, allowing for a fairer resource allocation. Secondly, {\it GGML} utilizes meta-learning to better perform beam shifting, thereby mitigating sidelobe interference among users and achieving higher spectral efficiency.

\begin{figure}[t]\vspace{-0mm}
	\begin{center}		\centerline{\includegraphics[width=0.42\textwidth]{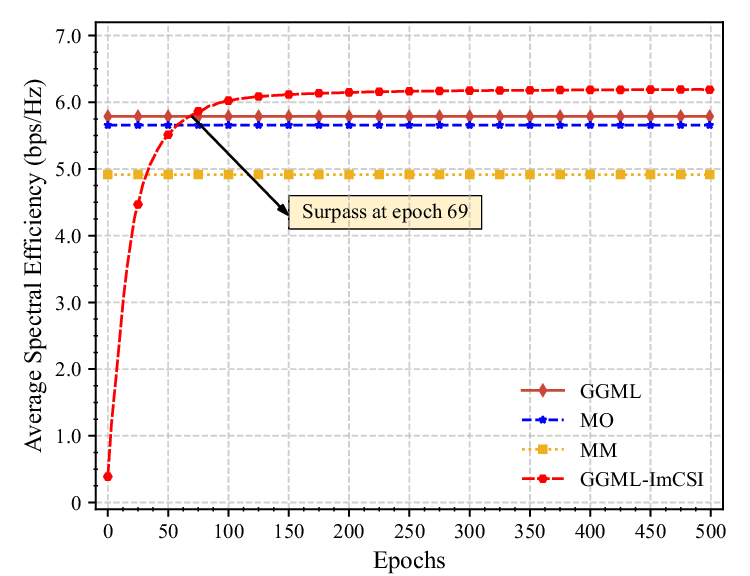}}  \vspace{-0mm}
	    \captionsetup{font=footnotesize, name={Fig.}, labelsep=period}  
        \caption[t]{\raggedright Average spectral efficiency versus the epochs, with a fixed channel error parameter of $\delta=0.2$.}
		\label{fig:convergence_ImCSI} \vspace{-8mm}
	\end{center}
\end{figure}

\subsection{Imperfect CSI}
This subsection evaluates the performance of the proposed {\it GGML-ImCSI} algorithm under imperfect CSI conditions. For simplicity, we assume equal channel uncertainty for all users, \ie, $\delta_{1}=\delta_{2}=\cdots=\delta_{K}=\delta$.
\par
We begin by exploring the convergence of the proposed algorithm under imperfect CSI conditions over \(10^2\) independent channel realizations, as shown in Fig. \ref{fig:convergence_ImCSI}. With CSI error parameter set at \(\delta=0.2\), {\it GGML-ImCSI} surpasses the {\it GGML} at the \(69\)-th epoch and can converge within 200 epochs, illustrating swift convergence due to the efficient search capabilities within the optimization landscape. The results also indicate that, the non-expanded {\it GGML} shows strong robustness to imperfect CSI when compared to conventional optimization algorithms. This increased resilience can be attributed to the adaptive capabilities inherent in meta-learning paradigms.
\begin{figure}[ht!]
	\centering
	\begin{subfigure}[b]{0.42\textwidth}
		\includegraphics[width=\textwidth, clip]{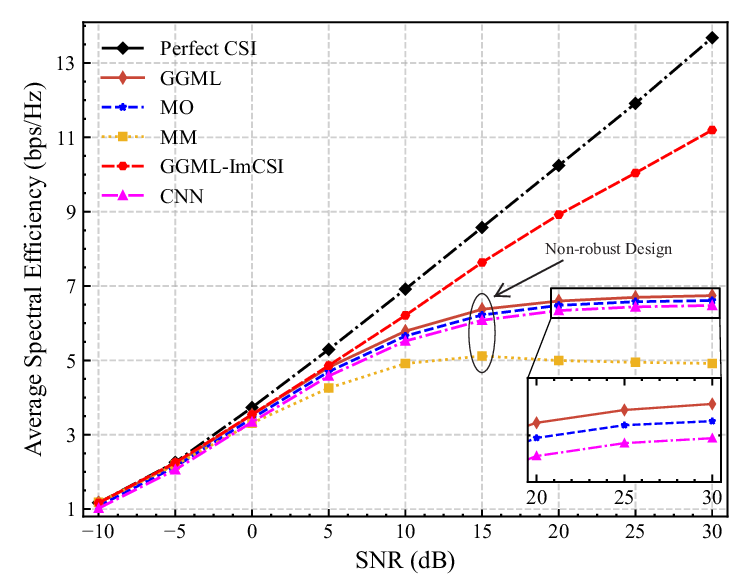}
		\caption{Average spectral efficiency performance versus SNR.}
		\label{fig_imcsi_a}
	\end{subfigure}
	\hspace{8mm} 
	\begin{subfigure}[b]{0.42\textwidth}
		\includegraphics[width=\textwidth, clip]{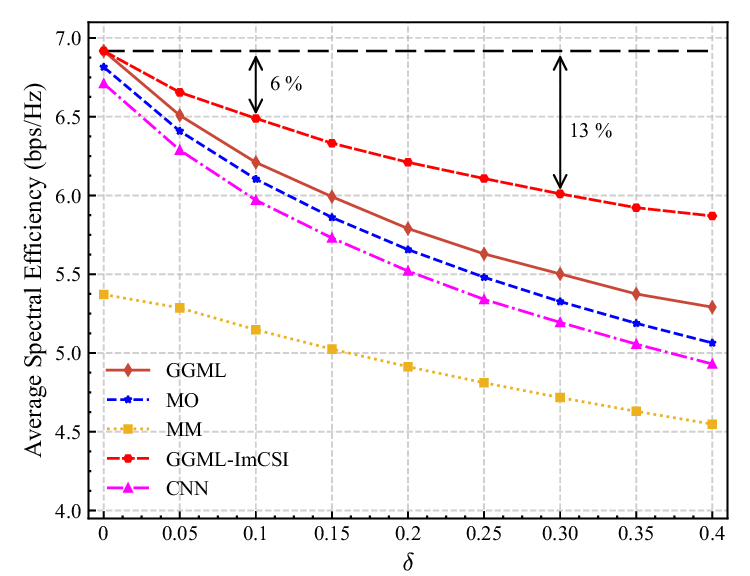}
		\caption{Average spectral efficiency performance versus the channel error parameter, $\delta$.}
		\label{fig_imcsi_b}
	\end{subfigure}
	\captionsetup{font=footnotesize, name={Fig.}, labelsep=period} 
	\caption{\raggedright Average spectral efficiency performance with imperfect CSI.}
	\label{fig:ImCSI_Performance}
\end{figure}
\par
Furthermore, we conduct simulations at distinct SNR and various levels of CSI error, represented by the variable $\delta$ in \eqref{channel_error}, to test the performance of different algorithms with imperfect CSI. Using the same setups in \ref{subsec:performance}, we present the average SE against SNR with imperfect CSI in Fig. \ref{fig_imcsi_a}. From this figure, we can observe that, {\it GGML-ImCSI} demonstrates enhanced robustness to imperfect CSI, particularly in the high SNR regions. Besides, in the absence of robust design, performance gain tends to plateau or even degrade in high-SNR regions. We explain this phenomenon in detail as follows.
\par
Intuitively, at low SNR, spectral efficiency is mainly constrained by the receiver noise, rendering the impact of CSI error almost negligible. However, in high-SNR regions, CSI error parameter gradually supersedes noise as the dominant factor, underscoring the importance of robust design. Specifically, as illustrated in Fig. \ref{fig_imcsi_a}, the performance gap between robust and non-robust designs widens as SNR increases.
\par
Then, we plot the average spectral efficiency against the channel error parameter \(\delta\) in Fig. \ref{fig_imcsi_b}. The performance loss grows with the increasing of \(\delta\), which is a predictable outcome. Specifically, for {\it GGML-ImCSI}, compared to perfect CSI (\ie, \(\delta=0\)), the system experiences approximately a 6\% loss in performance when the error power is 10\% of the channel gain (\ie, \(\delta=0.1\)), and a loss of 13\% when \(\delta=0.3\). In contrast, for the scheme without robust design, the corresponding losses are 13\% and 20\%. Therefore, the proposed scheme demonstrates strong robustness in the presence of imperfect CSI. Notably, as $\delta$ increases, the performance gap between {\it GGML-ImCSI} and {\it GGML} widens, indicating that the extended algorithm could improve performance under suboptimal channel estimation or rapidly fluctuating channels conditions.

%% file: conclusion.tex
\section{Conclusion}\label{sec:conclusion}
In this paper, we have developed an efficient hybrid precoding technique named GGML for mmWave downlink MU-MISO systems. To jointly optimize the digital and analog precoders, GGML integrated meta-learning and KAN, then optimized sub-problems using gradient descent from a local perspective, while updating the built-in network parameters from a global perspective. GGML replaced the native channel matrix with raw gradient information as network input, thereby further improving performance and accelerating the rate of convergence. 
\par
We have also extended our proposed algorithm to address scenarios with imperfect CSI. By rephrasing the maximin problem, we use GGML-ImCSI to solve the equivalent perfect CSI hybrid precoding design problem. Our simulation results demonstrated that the proposed algorithm had fast convergence and low computational complexity. Moreover, our algorithm outperformed the baseline method in various system configurations, and could even surpass fully digital WMMSE precoding in scenarios with a small number of RF chains. Even in imperfect CSI scenarios, our proposed method could counteract the negative impact of channel estimation errors, demonstrating its potential for practical deployment.
\par
The proposed algorithm is well-suited for MU-MIMO systems with hybrid architectures, as it only requires deploying lightweight neural networks at the user end, setting it apart from traditional neural network methods. Moreover, this approach will bring many open problems, such as optimizing convergence rate to meet the demands of highly dynamic environments, which are also left for future works.

%% file: appendix.tex
\appendix
\vspace{1mm}
\subsection{Proof of Proposition 2}\label{appendix1}
Let symmetric matrix $\mathbf{Q}_1 \succeq \mathbf{Q}_2 \succeq 0$ and define 
\begin{equation}
    \label{gt}
    g(t)={\Big(1+\mathbf{x}_k^H\mathbf{Q}(t){\big(\mathbf{T}_k\mathbf{Q}(t)+\sigma_k^2\mathbf{I}_N\big)}^{-1}\mathbf{x}_k\Big)}^{-1},
\end{equation}
where $\mathbf{Q}\left(t\right)=t\mathbf{Q}_1+(1-t)\mathbf{Q}_2 \; (0 \leq t \leq 1)$ and $\mathbf{x}_k=\mathbf{F}\mathbf{d}_k$.
\par
Then, based on the result in \cite{petersen2008matrix}, \ie, $\nabla\mathbf{X}^{-1}=-\mathbf{X}^{-1}(\nabla\mathbf{X})\mathbf{X}^{-1}$, we calculate the partial derivative of $g(t)$ with respect to $t$, and obtain \eqref{eq_partial} at the top of the page.
\begin{figure*}[t] 
\centering

    \begin{equation}\label{eq_partial}
        \frac{\partial g(t)}{\partial t} = - \frac{\mathbf{x}_k^H\Big(\mathbf{I}_N-\mathbf{Q}(t){\Big(\mathbf{T}_k\mathbf{Q}(t)+\sigma_k^2\mathbf{I}_N\Big)}^{-1}\mathbf{T}_k\Big)\Big(\mathbf{Q}_1-\mathbf{Q}_2\Big){\Big(\mathbf{T}_k\mathbf{Q}(t)+\sigma_k^2\mathbf{I}_N\Big)}^{-1}\mathbf{x}_k}{{\Big(1+\mathbf{x}_k^H\mathbf{Q}(t){\Big(\mathbf{T}_k\mathbf{Q}(t)+\sigma_k^2\mathbf{I}_N\Big)}^{-1}\mathbf{x}_k\Big)}^2}
    \end{equation}
\hrulefill
\end{figure*}
\par
Furthermore, applying the Kailath variant property and the
Searle’s identity ${(\mathbf{I}+\mathbf{AB})}^{-\mathbf{1}}=\mathbf{I}-\mathbf{A}{(\mathbf{I}+\mathbf{BA})}^{-\mathbf{1}}\mathbf{B}$, the partial derivative of $g(t)$ can be rewritten as 
\begin{equation}
    \label{eq_partial_2}
    \frac{\partial g(t)}{\partial t} = \omega \mathbf{y}_k^H\Big(\mathbf{Q}_1-\mathbf{Q}_2\Big)\mathbf{y}_k~\leq~0,
\end{equation}
where $\omega=-{\sigma_k^2}\Big/{{\Big(1+\mathbf{x}_k^H\mathbf{Q}(t){\Big(\mathbf{T}_k\mathbf{Q}(t)+\sigma_k^2\mathbf{I}_N\Big)}^{-1}\mathbf{x}_k\Big)}^2}$ is negative and $\mathbf{y}_k={(\mathbf{T}_k\mathbf{Q}(t)+\sigma_k^2\mathbf{I}_N)}^{-1}\mathbf{x}_k$.
\par
Hence, $g$ is decreasing in $ 0 \leq t \leq 1$. Subsequently, we can conclude that on the set of positive definite matrices $\mathbf{R}_{h_k}$, the function $e_k^{mmse}$ is decreasing. 
\par
Based on this, since $\mathbf{R}_{h_k}\succeq\mathbf{R}_{h_k}^{WC}$, $e_k^{mmse}$ can achieve an upper bound at the point $\mathbf{R}_{h_k}^{WC}$. Thus the proof of Proposition \ref{mmse_gradient} is completed.
\vspace{2mm}